\documentstyle[12pt,psfig]{article}
\def\issue(#1,#2,#3){{\bf #1}, (#2) #3} % AIP format

\def\APP(#1,#2,#3){Acta Phys.\ Polon.\ \issue(#1,#2,#3)}
\def\ARNPS(#1,#2,#3){Ann.\ Rev.\ Nucl.\ Part.\ Sci.\ \issue(#1,#2,#3)}
\def\CPC(#1,#2,#3){Comp.\ Phys.\ Comm.\ \issue(#1,#2,#3)}
\def\CIP(#1,#2,#3){Comput.\ Phys.\ \issue(#1,#2,#3)}
\def\CJP(#1,#2,#3){Chin.\ J.\ Phys. (Taipei)\ \issue(#1,#2,#3)}
\def\EPJC(#1,#2,#3){Eur.\ Phys.\ J.\ C\ \issue(#1,#2,#3)}
\def\EPJD(#1,#2,#3){Eur.\ Phys.\ J. Direct\ C\ \issue(#1,#2,#3)}
\def\IEEETNS(#1,#2,#3){IEEE Trans.\ Nucl.\ Sci.\ \issue(#1,#2,#3)}
\def\JHEP(#1,#2,#3){JHEP\ \issue(#1,#2,#3)}
\def\MPL(#1,#2,#3){Mod.\ Phys.\ Lett.\ \issue(#1,#2,#3)}
\def\NP(#1,#2,#3){Nucl.\ Phys.\ \issue(#1,#2,#3)}
\def\NIM(#1,#2,#3){Nucl.\ Instrum.\ Meth.\ \issue(#1,#2,#3)}
\def\PL(#1,#2,#3){Phys.\ Lett.\ \issue(#1,#2,#3)}
\def\PRD(#1,#2,#3){Phys.\ Rev.\ D \issue(#1,#2,#3)}
\def\PRL(#1,#2,#3){Phys.\ Rev.\ Lett.\ \issue(#1,#2,#3)}
\def\SJNP(#1,#2,#3){Sov.\ J. Nucl.\ Phys.\ \issue(#1,#2,#3)}
\def\ZPC(#1,#2,#3){Zeit.\ Phys.\ C \issue(#1,#2,#3)}

\def\be {\begin{equation}}
\def\ee {\end{equation}}
\def\bea {\begin{eqnarray}}
\def\eea {\end{eqnarray}}

\def\bc {\begin{center}}
\def\ec {\end{center}}

\def\lapp{\mathrel{\rlap{\raise.5ex\hbox{$<$}} {\lower.5ex\hbox{$\sim$}}}}
\def\gapp{\mathrel{\rlap{\raise.5ex\hbox{$>$}} {\lower.5ex\hbox{$\sim$}}}}
\def\r {\rightarrow}

\def \CH{{\tilde\chi}^{\pm}}
\def \LSP{\tilde\chi_1^0}
\def \PI{{\pi}^{\pm}}

\def \MLSP{m_{{\tilde\chi_1}^0}}
\def \MCH{m_{{\tilde\chi}^{\pm}}}
\def \MSNU{m_{{\tilde\nu}}}
\def \MPI{m_{{\pi}^\pm}}
\def \MCHMIN {\MCH^{min}}

\def \DM{\Delta m}
\def \MLL{m_{\tilde{l}_L}}

\begin{document}
\setcounter{page}{0}
\thispagestyle{empty}
\begin{center}
{\large\bf  Testing the AMSB Model via $e^+e^-\r\gamma{\tilde\chi}^{
+}{\tilde\chi}^{-}$}\\
\bigskip
{\normalsize Amitava Datta{\footnote {adatta@juphys.ernet.in}
\footnote{On leave of absence from Jadavpur University}}}\\
{\footnotesize
Department of Physics, Visva-Bharati, Santiniketan - 731 235,
India.}\\
{\normalsize  Shyamapada Maity{\footnote {shyama@juphys.ernet.in}}}\\
{\footnotesize
Department of Physics, Tamralipta Mahavidyalaya,\\
 Tamluk - 721636, West Bengal, India.}\\ 
\end{center}

\vskip 5pt

\begin{center}
{\bf ABSTRACT }
\end{center}
{\footnotesize  The possibility of detecting the signature of a
nearly invisible charged wino ($\CH$) decaying into a soft pion
 and the LSP($\LSP$),
predicted by the Anomaly Mediated Symmetry Breaking model, via the
process $e^+e^-\r\gamma{\tilde\chi}^{+}{\tilde\chi}^{-}$ at the Next
Linear Collider has been explored. Using the recently, proposed bounds
on slepton and wino masses derived from the condition of stability of the
electroweak symmetry breaking vacuum and employing some standared
kinematical cuts to supress the background, we find that almost
the whole of the allowed parameter
space with the slepton mass less than 1 TeV,  can be probed
at $\sqrt{s}$ = 500 GeV. Determination of the slepton and the chargino
masses from this signal is a distinct possiblity. Any violation of the
above mass bound will suggest that the standard vacuum is unstable and
we are living in a false vacuum.}

\vskip 10pt
\newpage

Supersymmetry (SUSY) has now been widely accepted as an elegant
alternative to the standard model(SM).  The current experimental
limits on the masses of the superpartners show that 
, even if they exist, they must be significantly heavier than
the corresponding SM particles. Thus SUSY must be broken. The
 mechanism of this  breaking is still unknown although several
interesting models have been proposed. Without such a specific
SUSY breaking mechanism,  the
most general minimal supersymmetric extension of the standard model
( MSSM) contains a
large number of soft breaking (SB) terms. The resulting model with
a large number of unknown parameters is, however, not very predictive.
Thus one attempts to construct a constrained MSSM with additional
theoretical assumptions on SUSY breaking.

In the minimal supergravity (mSUGRA) type models \cite{sugra,sugrarev}
one assumes that all the SM particles and their superpartners
belong to  the observable sector (OS). SUSY breaking takes place in
 hidden sector (HS), whose fields are all singlets under the SM
gauge group and are very heavy.
A contact interaction between the HS and the OS fields is introuced in the
K$\ddot{a}$hler potential by gravitational interactions which is  suppressed
 by the
Planck mass squared. This tree-level interaction induces SUSY breaking in
the OS; in such  models the gravitino mass is of the order of 1 TeV.
These models with the additional constraint of
radiative electroweak symmetry breaking \cite{ewb}
have only five more free parameters compared to the SM.

Recently, it has been pointed out  that if the OS and the
HS fields belong to  two distinct 3-branes separated by a finite distance
along a fifth compactified dimension, the above mechanism for
 transmitting  SUSY breaking from the HS to the OS fails.
However, a superconformal anomaly may induce the SUSY breaking in
the OS. Such soft breaking  terms  are also present in  SUGRA type models
, but they are suppressed in comparison to the usual soft-breaking
terms. To generate the weak scale masses of the
sparticles, the gravitino mass must be of the order of tens of TeV. Such
models are generically known as anomaly-mediated SUSY breaking (AMSB)
models \cite{amsb,amsb2}.

AMSB, alongwith the radiative electroweak symmetry breaking condition,
 fixes  the sparticle spectrum completely in terms of three parameters:
$m_{3/2}$ ( the gravitino mass ), $\tan\beta$ (the ratio of the vacuum
expectation values (VEV) of the two Higgs fields), and $sign(\mu)$.
The gaugino masses $M_1$, $M_2$ and $M_3$, and the trilinear
couplings (generically denoted by $A$) can be obtained from the
relevant renormalization group (RG)
$\beta$-functions and anomalous dimensions. The sfermion masses, as well
as the Higgs mass parameters, are also determined by $m_{3/2}$.
Unfortunately,
for sfermions that do not couple to asymptotically free gauge groups ({\em
i.e.}, the left and right sleptons,  $\tilde{l_L}$, $\tilde{l_R}$,
$l$= e, $\mu$, $\tau$), the masses come out to be tachyonic.
The remedy is sought by putting a positive definite mass squared term
$m_0^2$ in the GUT scale boundary conditions. Such
 terms can be justified  by appealing to the
presence of extra field(s) in the bulk. These models with a universal $m_0$
for all scalars are called the minimal AMSB (mAMSB) models
\cite{amsb,amsb2}.
The phenomenology of such models has been at the focus of attention of
many recent studies \cite{ggw,feng,amsb-pheno,probir,baer1,baer2},
and  our discussions will be restricted within the frame work of mAMSB models.

 One can determine the complete particle spectrum in terms of the above
free parameters. A crucial prediction very relevant for this paper
is as follows \cite{amsb,amsb2,ggw,feng,amsb-pheno}:
 the lighter chargino($\CH$) is almost degenerate with the lightest
neutralino($\LSP$), which we assume to be the lightest supersymmetric
particle (LSP). Both of them are heavily dominated by the wino component.

The near degeneracy  leads to
the most striking experimental signature of AMSB models, based on
the ``nearly invisible'' decay of the relatively long lived
$\CH$ to the LSP and a soft charged pion: $\CH\r\LSP\PI$ . 
If $M_2\geq$80GeV \cite{ggw}, then $\Delta m = \MCH -
\MLSP>  \MPI$. Consequently the  decay mode
$\CH\r\LSP\PI$ dominates over a large region of the allowed parameter
space(APS),
where the $\PI$ is rather soft for relatively small $\Delta m$. Thus
the chargino decays almost invisibly and the conventional search strategies
involving acoplanar leptons and/or jets + missing energy may not be applicable.

Due to the small $\DM$, however, the chargino has other distinguishing
characterestics which may be exploited in detecting them. For example,
they may have macroscopic decay lengths. This may lead to
heavily ionising tracks in the vertex detector without corresponding activities
in the calorimeter or in the muon chamber. Moreover the tracks end in soft 
pions
which may be observable, if the impact parameters
are clearly non zero and $p_T^{\pi}$ is sufficiently large.
These features may be utilised to identify $\CH$ production in the off-line
analysis if the event can be suitably triggered on.

There is a lower limit on $m_{3/2}$ coming from the  lower bound
$\MCHMIN = 86$ GeV \cite{chargino} on charginos decaying through the soft
pion mode from  direct searches at LEP.
This limit is roughly $m_{3/2}\sim$ 28-32 TeV. The detection of these charginos
are of crucial importance since they may happen to be the only sparticle( apart
from the LSP ) within the striking range of an early version of the Next Linear
Collider (NLC) at $\sqrt{s}$ = 500 GeV.

The search strategies for nearly invisible charginos were first studied by 
Chen et. al.
\cite{chen1,chen2} in the context of other models with such charginos.
 It was noted that for $\MPI<\DM<$230 MeV, the charginos may pass through 
several
layers of a  typical vertex detector leaving behind a heavily ionising track
 which by itself is a distinctive feature.
However, for $\DM$ in the upper half of the above range, they may  only
pass through one or two inner layers of the vertex detector
which may be difficult to distinguish from ,e.g, random hits due to detector
 noise. However, such charginos often end up in a soft
pion with an impact parameter(b) significantly larger than the impact parameter
resolution($b_{res}$) which is typically $\sim10^{-1}$cm \cite{chen2}.
If $\DM>$230 MeV the chargino tracks in the vertex detector
could be too short for proper identification.
However, the soft pions are usually more energetic in this case which may 
correspond
 to a better impact parameter resolution. Consequently the impact parameter
 of the
pion may be appreciably larger than $b_{res}$ inspite of the small chargino
decay length.

Any of the above features may suffice to identify nearly invisible charginos
in the off line analysis.
However, triggering based on activities in the vertex dector alone may be
problematic. Hence the process $e^+e^-\r\gamma{\tilde\chi}^{+}{\tilde\chi}^
{-}$       
was recommended, where the hard $\gamma$ and missing energy in the final state
can be easily triggered
on. The activities in the vertex detector and/or
soft pions may dramatically reduce the background from processes like 
$e^+e^-\r\gamma\nu\bar\nu$

The purpose of this note is to study the viability of this channel at NLC with
$\sqrt{s}$ = 500 GeV within the
framework of mAMSB models. We wish to emphasize that so far as this process
 is
concerned the mAMSB model is much more predictive and interesting than other
 models involving invisible charginos and radiative EW symmetry breaking.

The signal cross section is given by i) four $t$ channel sneutrino exchange
diagrams. ii) four $s$ channel Z-exchange diagrams  iii) four $s$ channel
$\gamma$ exchange diagrams \cite{chen1,chen2,shyama}.
In the most general case the production cross section depends on the
parameters $\mu$, tan$\beta$ and $M_2$ through the chargino mass and
mixing angles. These three parameters in addition  to the sneutrino mass
controls the size of the cross section. In the mAMSB model the lighter chargino
is a wino to a very good approximation. Thus the mixing angles ($\sim 1$)
 mildly depend on SUSY parameters.
As a result the cross sections is effectively a function
of two parameters only $\MCH$ and $\MSNU$. As was shown by Chen et. al. (to
be reviewed below), $\MCH$ can be determined from the kinematics alone. Thus
 the
size of the cross section may suffice to determine $\MSNU$, although this 
mass
could be well outside the kinematic reach of the collider,
which is quite possible for NLC at $\sqrt{s}$ = 500 GeV.

The cross section for the process $e^+e^-\r\gamma{\tilde\chi}^{+}{\tilde\chi
}^{-}$
was first computed  for models with $\MSNU\sim$1 TeV or larger
\cite{chen1,chen2}. In this
case the $t$ - channel $\tilde\nu$ exchange diagrams were justifiably  neglected.
However, for lower $\MSNU$ the cross section may be significantly smaller
\cite{shyama}. This happens due to
 destructive interferences between s-channel($\gamma$, $Z$ exchange) and t-
channel
diagrams.  The full cross section valid for all $\MSNU$ is given in \cite{shyama}.

     In the AMSB models the sleptons and sneutrinos may indeed be rather
light depending on
the choice of the common scalar mass $m_0$. This leaves open the possiblity
 that the
cross section could be significantly smaller than that obtained in the large
$m_0$ limit(see below for details). Fortunately as has been
pointed out recently, there is a lower bound on $m_0$ resulting in a corresponding
bound on $m_{\tilde{l_L}}$ and $\MSNU$ ($m_{\tilde{l_L}}\gapp$330 GeV
 most conservatively) \cite{abhijit}. This lower bound excludes the possiblity
of direct slepton pair production at NLC with $\sqrt{s}$ = 500 GeV. Thus 
$\CH$
and $\LSP$ are the only sparticles accessible to this accelerator.

The above bound arises by requiring that E-W symmetry breaking minimum of 
the scalar
potential be deeper than all possible charge/colour breaking minima \cite{ad}.
 Moreover,
thanks to the same constraints, there is an upper bound on the lighter chargino
mass for a given $m_0$(or slepton mass)(see table). The numbers in this table are
slightly different from the corresponding numbers in Table II of \cite{abhijit}.
This is due to the fact that we have taken into account loop induced electroweak
radiative correction \cite{ggw} to the chargino mass. The possible effects 
of
this correction were commented upon in \cite{abhijit} but they were not included
in the numerical results.

From the table it is clear that the lighter chargino mass is predicted to
 be
within the stricking range  of NLC at $\sqrt{s}$ = 500 GeV for a wide range of
slepton masses. Thus the signal of nearly invisible charginos is quite likely
to be seen. The absence of this signal on the other hand would
rule out a large part of the interesting parameter space of this model.

The above lower bound on the sparticles masses may be evaded by requiring that
the present SU(2)$\times$U(1) symmetry breaking minimum is a metastable false
vacuuam  which in principle can decay into much deeper true vacuum breaking
 charge
and colour symmetry. However, the decay time is larger than the age of the
universe \cite{hall}. In view of this interesting suggestion the measurment
of the slepton and the chargino mass as sketched above acquires
special significance. If the
mass indeed turn out to violate the bounds of \cite{abhijit} then that would
strongly indicate that we are living in a false vacuum.

In the mSUGRA models the violation of certain bounds on the $A$ parameter
\cite{ad} indicate the existence of a charge colour breaking vacuum.
Since this parameter is not directly related to any mass, this
experimental test is not straight forward. In contrast testing
the bounds on $\MCH$ or $\MLL$ is rather unambiguous from the
experimental point of view.

In Fig.1 we present the cross section at $\sqrt{s}$ = 500 GeV in the mAMSB
 model as a
function $\MSNU$ for $\MCH$ = 100, 150, 200 GeV.
From the figure it is seen that the cross section first decreases and then
increases with increasing of $\MSNU$. For small $\MSNU$ the $t$-channel diagrams
dominate and the effect of interference is relatively weak. For larger $\MSNU$
 interference effect is very prominent.
 This interference being destructive in nature, with the increasing of $\MSNU$
the cross section increases. It is clear from the figure that once $\MCH$ is
measured from the kinematics(see below) $m_{\tilde{l_L}}$, or $\MSNU$ may be
obtained from the size of the cross section.

The kinematical cuts used in computing the cross section are $p^{\gamma}_T\ge$10 GeV,
 $10^o\le\theta_\gamma\le170^o$.
These cuts are required to remove the radiative Bhabha and other backgrounds 
\cite{chen1}.
 To ensure the observability of the soft pions
, we further require that only one of the two soft pions satisfy
 $p^{\pi}_T\ge$200 MeV, $\eta_\pi\ge$2.5 \cite{probir}.

In the table we present the cross section with above kinematical cuts
on the $\gamma$ and the soft pion(see below). From the table it is clear
that the entire allowed range of $\MCH$ except for $m_0\approx$1 TeV, leads
to hundreds of background free events at NLC with $\sqrt{s}$ = 500 GeV
 and $\cal{L}$ = 50 $fb^{-1}$.

We next analyse the track length of the $\MCH$ and the inpact parameter of 
the soft
pion to check whether these characterestics are adequate to make the signal
background free.

Although $\sigma$ is largely insensitive to tan$\beta$ and sign$\mu$, the track 
length ($l$) of the chargino depends sensitively on these parameters, where
$l$=$\beta{\tau}_{lab}$.
This is due to the fact that $\DM$ and, hence, the $\CH$ life time in its
rest frame($\tau$) depends  quite a bit on these
parameters. We have studied the distributions of these two observables for
the entie
parameter space allowed by the bounds of \cite{abhijit}.

We find that at $\sqrt{s}$ = 500 GeV, $l\lapp$4 cm. This is due to the fact
that at low $M_2$ (i.e. small $\MCH$), $\DM$ is large \cite{ggw}. This lead
to a small $\tau$ and consequently a short track. On the other hand for large
$M_2$ although $\DM$ is favourable, $\beta$ at $\sqrt{s}$ = 500 GeV is not
enough to give a long track.

In \cite{chen2} the characteristics of a typical Silicon Vertex Detector(SVD)
has been quoted from a CDF report \cite{CDF}.
From the analysis of \cite{chen2} it is clearer that for $l\lapp$ 4 cm. a
 chargino is, not
likely to traverse more than a few inner layers of the vertex detector and 
this
characteristic may not be sufficiently distinctive for suppressing the background.
However, the estimates of \cite{chen2} are based on the CDF-II detector. It
is quite possible that the vertex detector at NLC will have layers closer to
 the beam pipe. The fact that the beam at NLC will be narrower than that
at Tevatron strengthens this expectation. If that is the case then even
$l\lapp$ 4 cm. may help to reduce the background.
 The impact parameter of the pion is, however,
large enough for the bulk of the APS and helps to reduce the background even if
$l$ is small. Assuming an impact parameter resolution
$b_{res}\sim$0.1 cm \cite{chen2}, we have required $b>$0.5 cm as the criterian
for the detectability of the pion. In fact $b_{res}$ could be significantly
smaller \cite{chen2} depending on the pion energy. Our assumption is, therefore,
quite conservative.

 For all $m_0$ in the table, except for $m_0$=
1 TeV, the number of events with at least one detectable soft pion exceeds 
10 as
long as $\MCH$ is below the upper bound $(\MCH)_{max}$.
For $m_0$ = 1000 GeV and tan$\beta$ = 5 a $\CH$ with mass close to the upper
 limit  lies outside the kinematic reach
of NLC with $\sqrt{s}$ = 500 GeV(see table). The search limit in this case
is found to be $\MCH$= 219 GeV. The corresponding $b$ distributions
of the pion is shown
in fig-2. Requiring $b>$0.5 cm, fig-2 corresponds to 10 events.

It is quite possible that
at higher $\sqrt{s}$ the chargino track length will be  significantly larger.
This is illustrated in Fig-3. We have taken $\MCH$=262 which corresponds 
to the
upper bound on $\MCH$ for $m_0$= 1 TeV. and $\sqrt{s}$ = 1 TeV. According
to \cite{chen2} for $\beta{\tau}_{lab}\gapp$7 cm, the charginos will pass
through at least four layers of the SVD. Thus for entire allowed range
of the chargino masses corresponding to $m_0$=1 TeV, glimpses of
heavily ionising chargino tracks can be seen in the SVD at $\sqrt{s}$= 1 TeV.

We next discuss the determination of [$\MCH$] from the observable
$m_{Z^\star}\equiv([p^{e^+}+p^{e^-}-p^{\gamma}]^{1/2})/2 $ \cite{chen1,chen2}.
It follows that $m_{Z^\star}\gapp\MCH$ for the signal. The
results for $\sqrt{s}$ = 500 GeV are shown in fig-4
for  $\MCH$= 100, 150, 200. Although the
kinematical cuts distort the lower edge of the distribution to some
extent, $\MCH$ can be measured with an accuracy of a few percent
except for $\MCH\gapp$ 200 GeV.

Once $\MCH$ is determined $\MSNU$ can be measured to a good approximation
from the size of the cross section.  Even if $\MSNU$ is determined approximately,
this information, with some luck, may be utilised in distinguishing the mAMSB
model from other competing models with small $\DM$.
For example in the string motivated model of \cite{chen1,chen2} the slepton
mass is necessarily large ($\sim$ 1 TeV) and as discussed above the cross
section for the same $\MCH$ is expected to be larger.

This illustrates the advantage of this signal over a similar discovery channel,
$p \bar{p}\r{\tilde\chi}^{+}{\tilde\chi}^{-} g$, at hadron colliders
(J. Feng {\em et. al.} in \cite{amsb-pheno}).
 In the later
case neither $\MCH$ nor the slepton mass can be reliably measured from the
observed signal. Thus the underlying model may not be identified.

The model underlying the nearly invisible charginos may be unearthed by
exploiting the advent of polarised electron beams. For example, with a
right polarised beam for which the $t$ - channel diagrams decouple, the
cross section will be significantly larger than the unpolarised cross
section if the slepton masses happen to be small. In contrast in the string
motivated model with heavy sleptons cross sections are not likely to show
any drastic change. The computation of the signal for polarised beams
is under progress.

In conclusion we reiterate that the signal
$e^+e^-\r\gamma{\tilde\chi}^{+}{\tilde\chi}^{-}$ looks very promising
in the mAMSB model. Most interestingly there is i) a lower bound on the
slepton mass ($m_{\tilde{l}_L}\ge$ 330 GeV) and ii) an upper bound on
chargino mass(see table) for a given slepton mass \cite{abhijit} which
makes this model very predictive. In view of these bounds  $\CH$ and
$\LSP$ are the only sparticles within the striking range of NLC at
$\sqrt{s}$ = 500 GeV.

For a wide range of common scalar mass ($m_0\lapp$1 TeV) the signal can
 be detected by exploiting
the soft pions from $\CH$ decay for the entire allowed range of $\MCH$.
For $m_0$ = 1 TeV, $\MCH\le$219 can be detected at $\sqrt{s}$= 500 GeV.
Assuming the charcteristics of typical SVD \cite{chen2} it appears that the
chargino tracks in the SVD may not be long enough to be observable
at $\sqrt{s}$ = 500 GeV. However, for $\sqrt{s}$ = 1 TeV  chargino
tracks with sufficient length which end in a soft pion can be probed in SVD.

The measurment of $\MCH$ and $\MSNU$ from the $m_{Z^\star}$ distribution
and from the total cross section has been discussed.
This measurments would be of crucial importance because if
the bounds of \cite{abhijit} are violated, that
would strongly indicated that we are living in a false vacuum \cite{hall}.
Further this mass determination, with some luck, may reveal the model underlying
these invisible charginos.

The muon anomalous magnetic moment has
recently been measured by E821 Collaboration at Brookhaven
\cite{muon} with an unprecedented precission.
Apparently there is a 2.6 $\sigma$  discrepancy between the experimental
result and the SM  prediction( for discussions and references to earlier 
works see, e.g.,
\cite{muon, marciano}). However, this statement is based on the assumption 
that the computation of the hadronic
corrections to the  muon anomalous momnent ,
which is the largest source of uncertainty in the SM  prediction, is
under control. There is already a claim that if these corrections are 
computed in a way different from that in \cite{davier}, 
which \cite{muon,marciano} quote, then there may be
agreement between the SM and the data \cite{yndu}. Thus ruling out the SM 
or some specific extension of it
on the basis of the alleged discrepancy  requires a cautious approach.

If, however, the SM prediction is accepted at its face value, then one of 
the  recent analyses indicates that the mAMSB model is disfavoured by the data
\cite{moroi}. On the contrary
it has been shown in \cite{tata} that while this model with low tan$\beta$
  is indeed disfavoured,
the one with large  tan$\beta$ is allowed. A critical reexamination of 
these claims are beyond the scope of this paper.

It may be noted that the bounds  of \cite{abhijit}
are quite restrictive for large   tan$\beta$ and the signal
 cross section is observable for large as well as
small values of  this parameter ( see the table).
 At the moment, therefore,  the model and a large part of
the APS scanned in this paper has no obvious conflict
with the data on muon anomalous magnetic  moment.

\vskip 1pt
\begin{table}
\begin{center}
\begin{tabular}{|c|c|c|c|c|c|}
\hline
\hline
% & & & & \\
tan$\beta$ &sign of&$m_0$&$m_{3/2}$&$(m_{{\tilde\chi}^{\pm}})_{max}$&$\sigma$ \\
&$\mu$ & & & &in $fb$ \\
\hline
5&+&500&35.8&123.12&22.35\\
5&-&500&35.7&116.23&29.30\\
5&+&700&51.6&177.53&5.71\\
5&-&700&51.5&173.56&13.19\\
5&+&1000&76.0&262.96&-\\
5&-&1000&75.9&261.61&-\\
35&+&500&33.3&111.45&28.46\\
35&-&500&33.5&110.85&30.16\\
35&+&700&48.2&163.99&13.04\\
35&-&700&48.5&164.21&14.52\\
35&+&1000&71.1&245.25&-\\
35&-&1000&71.5&246.19&-\\
\hline
\hline
\end{tabular}
\vskip .5in
\end{center}
{\bf Table}: {\it Upper bounds on $\MCH$ and corresponding $\sigma_{{\tilde
\chi}^{+}
{\tilde\chi}^{-}\gamma}$ for given values of $m_0$ at $\sqrt{s}$ = 500 GeV. 
The kinematical cuts are discussed in the text.}
\end{table}
\vskip 1pt
\vspace*{-1.in}
\hspace*{-1.1in}
{\hbox{
\psfig{figure=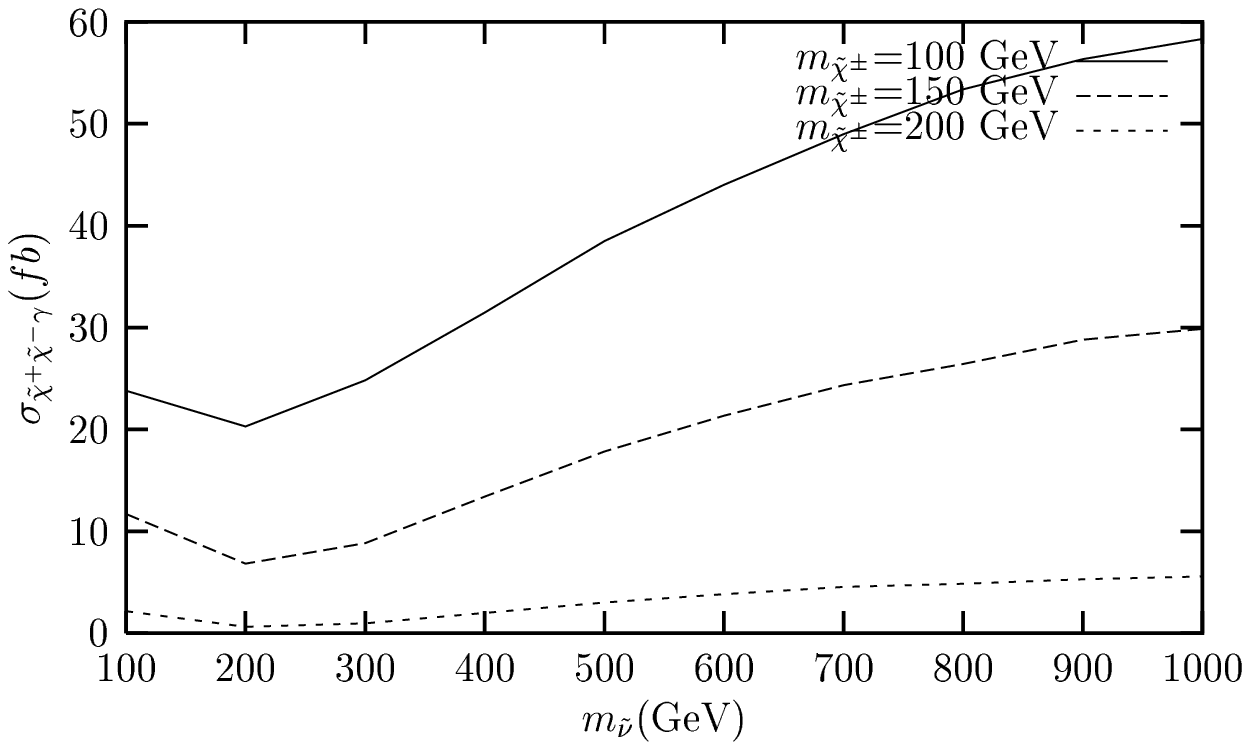,height=9in,width=7in}
}}
\vspace*{-5.in}
\begin{center}
{\bf Fig. 1}: {\it The variation of $\sigma_{{\tilde\chi}^{+}{\tilde\chi}^{-}\gamma}$
 with $m_{\tilde{\nu}}$ for three values of $\MCH$ at $\sqrt{s}$
= 500 GeV}.
\end{center}
\vspace*{.5in}

\vspace*{-1.in}
\hspace*{-1.1in}
{\hbox{
\psfig{figure=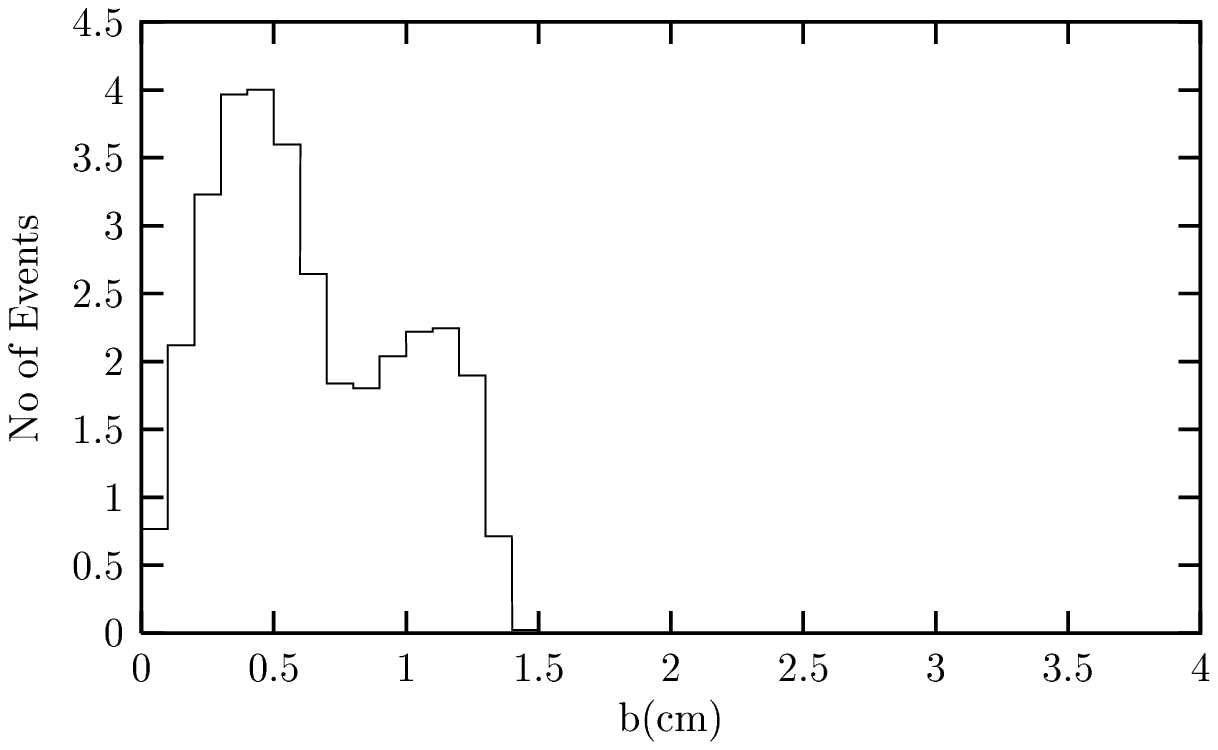,height=9in,width=7in}
}}
\vspace*{-5.in}
\begin{center}
{\bf Fig. 2}: {\it Impact Parameter distribution for $\MCH$ = 219 GeV at =
$\sqrt{s}$ 
500 GeV.}
\end{center}
\vspace*{.5in}

\vspace*{-1.in}
\hspace*{-1.1in}
{\hbox{
\psfig{figure=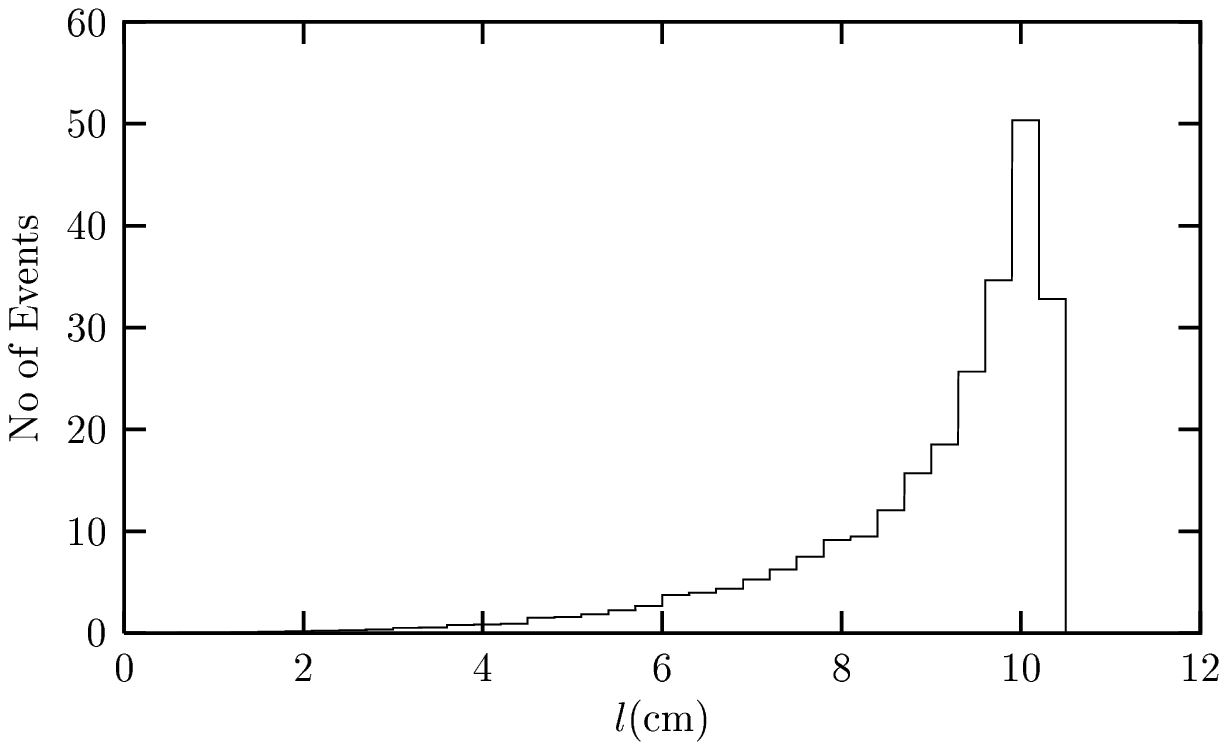,height=9in,width=7in}
}}
\vspace*{-5.in}
\begin{center}
{\bf Fig. 3}: {\it Decay Length distribution for $\MCH$ = 263 GeV at 
$\sqrt{s}$ = 1000 GeV}
\end{center}
\vspace*{.5in}

\vspace*{-1.in}
\hspace*{-1.1in}
{\hbox{
\psfig{figure=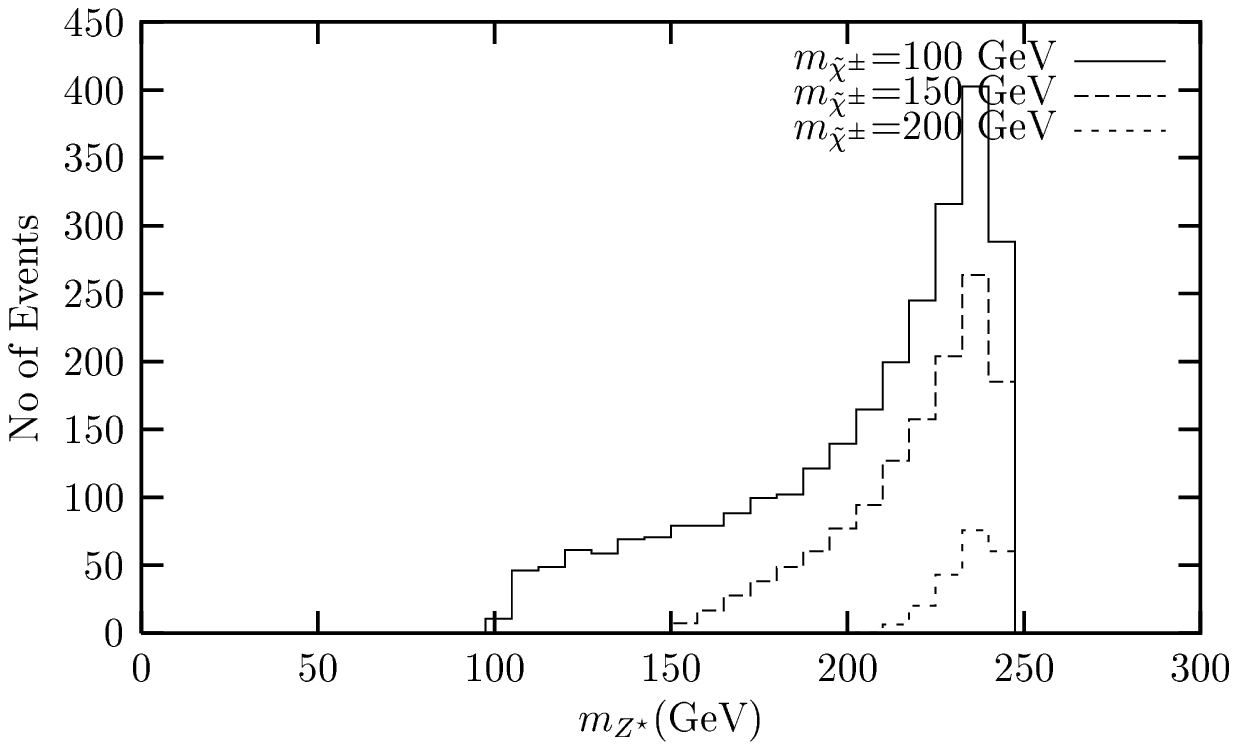,height=9in,width=7in}
}}
\vspace*{-5.in}
\begin{center}
{\bf Fig. 4}: {\it $m_{Z^\star}$ distribution for three values of $\MCH$ at
 $\sqrt{s}$ = 500 GeV}
\end{center}
\vskip 1pt
{\em Acknowledgements:}\\
The work of AD was supported by DST, India(Project No. SP/S2/K01/97)
and BRNS, India (Project No. 37/4/97 - R \& D II/474).

\end{document}